\documentstyle[prl,aps,twocolumn]{revtex}
\newcommand{\Tr}{{\rm Tr}}

\begin{document}

\draft
\tighten
\def\footnoterule{\kern-3pt \hrule width\hsize \kern3pt}


\title{On the uniqueness of the expected stress-energy tensor in
renormalizable field theories}

\author{L.L. Salcedo}

\address{
{~} \\
Departamento de F\'{\i}sica Moderna \\
Universidad de Granada \\
E-18071 Granada, Spain
}

\date{\today}
\maketitle

\thispagestyle{empty}

\begin{abstract}
It is argued that the ambiguity introduced by the renormalization in
the effective action of a four-dimensional renormalizable quantum field
theory is at most a local polynomial action of canonical dimension
four. The allowed ambiguity in the expected stress-energy tensor of a
massive scalar field is severely restricted by this fact.
\end{abstract}


\pacs{PACS numbers: 04.62.+v, 11.10.Gh}

In a recent paper\cite{Tichy:1998ws}, it has been rightly noted that
Wald axioms\cite{Wald:1994} allow for a much larger ambiguity in the
expectation value of the renormalized stress-energy tensor for massive
scalar particles than for massless ones. It was previously assumed
\cite{Wald:1994} that, as in the massless case \cite{Horowitz:1980fj},
in the massive case the Wald axioms were also sufficient to completely
determine the expected stress-energy tensor as a functional of the
state and background metric (up to the usual two parameter ambiguity)
without the need of further physical input. The paper
\cite{Tichy:1998ws} clears up this misconception and so it is relevant
to the foundations of the point-splitting procedure in the massive
case (see also \cite{Moretti:1999rf}).

In \cite{Tichy:1998ws}, the expected stress-energy tensor was computed
for minimally coupled massive scalar fields in a nearly flat spacetime
using two different formalisms, but no enlargement in the ambiguity
was actually found as compared to the massless case. In this comment,
we want to note that this result should not be surprising since,
besides Wald axioms, there are additional criteria which can be used
to drastically reduce the ambiguity.

The problem of the ultraviolet ambiguities in quantum field theory
appears in flat as well as in curved spacetime. It seems sensible to
assume that the ultraviolet sector will depend only on local
properties of the spacetime manifold and will not be affected by
global topological aspects \cite{note1}. Therefore we can restrict to
spacetimes which are topologically equivalent to R$^n$ (this certainly
covers the case of nearly flat spaces). The possibility of particle
creation induced by a local curved region or even a different in and
out vacuum due to different asymptotic metric tensors are not specific
of curved spacetimes since these phenomena can also occur in the
presence of suitable external fields in flat spacetime
\cite{Itzykson:1980}. Another issue is that of the appropriate measure
for the functional integration on the configuration space of the field
to be quantized. General covariance requires it to depend on the
metric \cite{Toms:1987sh}, however this is also not specific of curved
spaces; for instance, the axial anomaly for Dirac fermions implies
that the fermion measure should depend on the background vector and
axial field configurations \cite{Fujikawa:1979ay}. Therefore, for the
assumed spacetime topologies, we can regard the quantum field theory
as one on a flat spacetime with the metric $g_{\mu\nu}(x)$ playing the
role of a particular kind of background field and general covariance
as a particular symmetry of the action (under simultaneous
transformations of the quantum and background fields).

The problem of the appropriate functional measure is in fact
completely equivalent to that of the ultraviolet ambiguities
introduced by the renormalization. On the one side, the measure is not
entirely well-defined because of the ultraviolet divergences and their
necessary renormalization, but on the other side the measure cannot be
completely arbitrary, since that would amount to end up with a
completely arbitrary action, unrelated to the original action of the
theory. A practical point of view is to use perturbation theory (or
other expansions) in order to isolate the ultraviolet divergent
pieces. All contributions which are finite without regularization are
naturally postulated to be free from ambiguities and all acceptable
renormalization prescriptions are required to reproduce those finite
values. On the other hand, contributions which are divergent under any
expansion are subject to renormalization by subtraction of appropriate
counterterms. Since finiteness does not fix the counterterms uniquely,
the renormalized contributions become finite but not unique. The
requirement of finiteness allows for very general
renormalizations. The class of allowed regularizations can naturally
be restricted by demanding that they should be independent on the
background fields and parameters of the theory, otherwise the
regularization would mask the dependence of the effective action and
expectation values on these background fields and parameters (e.g., we
require a cutoff not to change under variations of the mass of the
quantum field or the metric tensor).

For definiteness, let us consider a single quantum scalar field
$\phi(x)$ with Lagrangian density ${\cal L}(x)$. More general cases
are discussed below. In general ${\cal L}(x)$ will depend on some
parameters $\lambda_i$ (such as masses, coupling constants and
spectroscopic factors) and background fields, $L_i(x)$. Without loss
of generality we can include the parameters $\lambda_i$ into the set
of $L_i(x)$ since they can be seen as scalar background fields which
happen to take a constant configuration. Moreover, we can assume that
all background fields (including the parameters) couple linearly to local
operators depending on $\phi$ and its derivatives. For instance, the
Lagrangian density of a minimally coupled scalar field
\begin{equation}
{\cal L}(x)=
\frac{1}{2}\sqrt{-g}(g^{\mu\nu}\partial_\mu\phi\partial_\nu\phi-m^2\phi^2)
\label{eq:1}
\end{equation}
can be seen as a particular case of the more general theory 
\begin{equation}
{\cal L}(x)=
\frac{1}{2}H^{\mu\nu}\partial_\mu\phi\partial_\nu\phi-\frac{1}{2}G\phi^2
\label{eq:2}
\end{equation}
where $H^{\mu\nu}(x)$ and $G(x)$ are arbitrary independent external
fields; at the end of the calculation they can be particularized to
their values in eq.(\ref{eq:1}).

It is standard to describe the properties of the quantum field
$\phi(x)$ using the effective action $\Gamma[\phi,L]$
\cite{Brown:1980qq}. The effective action is a finite quantity from
which derive all renormalized Green functions, i.e., all expectation
values. It will be sufficient for us to use the quantity $W[L]$
defined as $\Gamma[\phi,L]$ evaluated at its minimum with respect to
$\phi$. This is equivalent to the logarithm of the partition
functional in the absence of external currents, $\int
[d\phi(x)]e^{iS[\phi]}$. The quantity $W[L]$ is also frequently
referred to as the effective action and provides the expectation
values of operators coupled to the external fields. For instance, the
expectation value of the stress-energy tensor $T_{\mu\nu}$ is given by
the functional variation of $W$ with respect to $g_{\mu\nu}$.
Conservation of $T_{\mu\nu}$ follows if $W$ is a scalar under general
coordinate transformations \cite{Weinberg:1972}. Let us remark that
the renormalization of the effective action introduces ambiguities in
this quantity but no further ambiguities appear in the extraction of
the expectation values since this is done through differentiation or
minimization (diagrammatically this only involves tree level graphs).

After these general considerations, let us come to the issue of the
allowed ambiguity introduced by the renormalization. We begin by
considering the scalar field without self-interaction in
eq.~(\ref{eq:2}). The Lagrangian density is quadratic and, after
integration by parts, it can be rewritten as ${\cal L}(x)=
\frac{1}{2}\phi A\phi$ where $A$ is a second order differential
operator depending on the external fields. As is well known, the
effective action is formally given by $W[H,G]=
\frac{i}{2}\Tr\log(A)$. This is a formal expression which is
ultraviolet divergent and so requires renormalization. To do
perturbation theory, the Lagrangian is separated as
\begin{equation}
{\cal L}(x)=
\frac{1}{2}\eta^{\mu\nu}\partial_\mu\phi\partial_\nu\phi-\frac{1}{2}m_0^2\phi^2+
\frac{1}{2}\delta H^{\mu\nu}\partial_\mu\phi\partial_\nu\phi-\frac{1}{2}\delta
G\phi^2
\end{equation}
Diagrammatically, $\frac{i}{2}\Tr\log(A)$ corresponds to one-loop
graphs where the field $\phi$ runs in the loop with mass $m_0^2$ and
with any number of insertions of the external fields $\delta
H^{\mu\nu}(x)$ and $\delta G(x)$. In order to find the dependence on
the field $\delta G$ of the allowable ambiguities, some standard
diagrammatic results can be applied \cite{Collins:1984}. All diagrams
with a sufficiently large number of insertions of $\delta G$
(depending on the spacetime dimension) are ultraviolet
convergent. This is because each new insertion implies a new
propagator line thereby decreasing the degree of divergence of the
graph. This implies that all pieces in the effective action with a
sufficiently large power of $\delta G$ are ultraviolet
finite. Therefore, the divergent part is a polynomial in the external
field $\delta G$. Likewise, taking derivatives with respect to the
external momenta carried by $\delta G(x)$ also increases the
convergence and eventually the graphs become convergent. This implies
that their divergent component is a polynomial in the external momenta
of $\delta G(x)$.  Equivalently, the divergent contribution to the
effective action functional contains a finite number of partial
derivatives of the $\delta G(x)$. (More precisely, it is a polynomial
in the operators $\partial_\mu$ acting on $\delta G$.)

The polynomial character of the dependence on $\delta G$ of the
divergences also follows easily from the formal expression
$\frac{i}{2}\Tr\log(A)$, without resorting to perturbation theory,
since it is clear that after a sufficient number of functional
derivatives with respect $\delta G$, there will be a sufficient number
of powers of $A$ in the denominator so that the expression becomes
fully ultraviolet convergent. The argument also applies to the
dependence on $m_0^2$, since taking a derivative with respect to the
mass also increases the convergence. (Note that this kind of arguments
assumes that the regularization does not depend on the parameters to
be varied.)

From straightforward power-counting\cite{Collins:1984} it follows
that, as a function of $m_0^2$ and $\delta G$, the divergent component
in the effective action functional is a local polynomial constructed
with $m_0^2$, $\delta G(x)$ and its derivatives, of canonical
dimension at most four (or more generally, of canonical dimension at
most $D$ in $D$-dimensional field theories). The canonical dimension
refers to that carried by $m_0^2$, and the field $\delta G(x)$ and its
derivatives. The total mass dimension (of the effective action density
ambiguity) is four and comes, in addition, from the derivatives of
$H^{\mu\nu}(x)$ and from possible further dimensionful parameters
(with non negative mass dimension, as we will see) introduced by the
renormalization procedure. The divergent components are canceled by
adding new terms to the action. After such a renormalization, the
effective action is finite, but not unique. The renormalization leaves
an ambiguity which is again a local polynomial in $m_0^2$ and $\delta
G(x)$ and its derivatives of canonical dimension at most
four. Therefore, two mathematically acceptable versions of the
renormalized effective action will differ at most by a local
polynomial of dimension four. This is necessary if they are to
coincide in their ultraviolet finite components. This latter
requirement is natural since such ultraviolet finite components can be
computed without any regularization and hence have a unique and
well-defined value.

The previous discussion implies that ambiguities of the form
considered in \cite{Tichy:1998ws}, such as very general functions of
$R/m^2$ ($R$ being the scalar curvature), although allowed by Wald
axioms, cannot actually appear. Only polynomials in $m^2$ may
appear. This is because taking $n$ derivatives with respect to $m^2$
in the effective action yields (the connected part of) the expectation
value of $\left(\int dx^4\sqrt{-g}\,\phi^2(x)\right)^n$, which is a
completely ultraviolet finite quantity when $n>2$ and hence free from
ambiguities. However, two versions of the effective actions differing
by an arbitrary function of $m^2$ would yield different result for
$\langle\left(\int dx^4\sqrt{-g}\,\phi^2(x)\right)^n\rangle_c$ and so
at most one these effective actions could be acceptable\cite{note3}.

As it stands, the unique requirement of having a consistent
ultraviolet finite component permits a rather large ambiguity in the
effective action.  This allows in particular adding terms containing
new parameters or even external fields not present in the original
Lagrangian. For instance, one can choose to renormalize the theory in
such a way that the effective action depends on $m_0^2$ and $\delta
G(x)$ as independent variables, although the Lagrangian was only a
function of $G(x)=m_0^2+\delta G(x)$, or introduce a mass scale even
if the underlying theory is massless \cite{note2}. Usually, the choice
of renormalization is restricted in order to preserve the symmetries
in the Lagrangian. When no such choice is available the symmetry is
anomalously broken by the quantization.

Next, let us analyze the dependence of the allowed ambiguities on the
field $H^{\mu\nu}(x)$. The difference with the previous case of the
field $\delta G$ is that $H^{\mu\nu}$ is coupled to an operator
containing two derivatives of $\phi(x)$. In the Feynman rules in
momentum space, this translates to two momenta for each insertion of
$H^{\mu\nu}$ and so, higher orders in $H^{\mu\nu}$ are no longer
necessarily ultraviolet convergent\cite{note4}. It is still true,
however, that taking a sufficient number of derivatives with respect
to the external momenta carried by $H^{\mu\nu}$ yields a ultraviolet
convergent integral, therefore, the allowed ambiguity in the effective
action will contain no more than four derivatives of $H^{\mu\nu}$
(more precisely, the ambiguity will be a polynomial, at most of degree
four, on the operator $\partial_\mu$ acting on $H^{\mu\nu}$).

Since covariance under general coordinate transformations is a
symmetry free from anomalies it is natural to renormalize the theory
imposing this symmetry. This restricts the class of effective actions
so that they are scalars under such transformations, and thus the
possible ambiguities within this restricted class of effective actions
are also scalars.  Consequently the ambiguity in the effective action
will depend on $H^{\mu\nu}(x)$ only through the Riemann tensor (as
follows, for instance, from using Riemann coordinates). More
precisely, the allowed ambiguities will be of the form $\sqrt{-g}$
times a function of the Riemann tensor and the scalar
$G(x)/\sqrt{-g}$, and their covariant derivatives. The dependence on
$G/\sqrt{-g}$ has already been discussed, and is a local polynomial of
dimension at most four. The dependence on the Riemann tensor is highly
restricted by the fact that there can be at most four derivatives of
the metric tensor\cite{note5}. For instance, for the massive scalar
field of eq.(\ref{eq:1}), and assuming that no new dimensional
parameters are introduced by the renormalization, the ambiguity will
contain only the terms \cite{Brown:1980qq} $m^4$, $m^2 R$, $R^2$,
$R_{\mu\nu}R^{\mu\nu}$ and the Gauss-Bonnet density (which being a
topological term does not contribute to the stress-energy expectation
value nor to the semiclassical field equations). More generally, if
the renormalization introduces new dimensionful parameters,
dimensional counting allows further terms of the form $M^4$, $m^2M^2$
and $M^2R$ \cite{Brown:1980qq}.

The previous considerations completely cover the case of scalar
fields without self-interaction. They can be extended to the case of
arbitrary renormalizable theories with scalars, spin-1/2 fermions and
gauge fields, as follows.

As is well-known, the effective action $W[L]$ comes from adding all
one-particle-irreducible graphs with any number of insertions of the
external fields and without external legs (that is, legs of the
quantum fields). The quantum fields run on the propagators
corresponding to internal lines in the graph. Each of these leg-less
Feynman graphs has a superficial degree of divergence as well as
possible subdivergences. It is a standard result of renormalization
theory\cite{Collins:1984} that the renormalization procedure is such that
all subdivergences have already been removed due to counterterms of
lower order than the graph under consideration. Thus, only the
superficial or overall divergence needs to be canceled. (This holds in
fact for all Feynman graphs, with or without legs.)

When the theory is perturbatively renormalizable there is only a
finite number of operators in the renormalized Lagrangian, coming from
operators in the bare Lagrangian plus possibly new operators
introduced as counterterms. (In fact, and up to symmetries, all local
polynomial operators of canonical dimension at most four will appear.)
The corresponding parameters, such as masses and coupling constants,
are not fixed by the renormalization procedure but in principle they
can be fixed to their experimental values. Thus we will not count them
for the possible ambiguity in the effective action. (Note that for
case considered before of quantum fields without self-interaction
these parameters need not be renormalized.)

Thus when computing the leg-less graphs of the effective action $W[L]$
in a renormalized theory, it is only necessary to cancel the overall
divergence. Again, power-counting arguments show that for a
renormalizable theory the corresponding counterterms are local
polynomial operators of canonical dimension at most four, constructed
with the external fields and their derivatives. This holds for all
external fields except $g_{\mu\nu}(x)$ because the metric tensor
couples to the kinetic energy operator (and so new insertions of
$g_{\mu\nu}(x)$ do not increase the convergence). The dependence on
the metric tensor is restricted by general covariance, which requires
a dependence only through the Riemann tensor and its covariant
derivatives and with canonical dimension at most four.

For the effective action $\Gamma[\phi,L]$, the previous arguments hold
as well. The effective action comes from graphs which may contain
external legs of the quantum fields. After the theory has been
renormalized only the overall divergence needs to be cured and for
renormalizable theories this introduces terms which are local
polynomials and of canonical dimension at most four, constructed with
the quantum fields, the external fields (or the Riemann tensor) and
their covariant derivatives.

Two final comments are in order. First, in self-interacting theories
there can be ambiguities of purely non-perturbative origin which would
not leave a trace at the perturbative level and so they are not
covered by the present analysis. This is the case of the renormalon
ambiguities introduced by Borel resummation of the perturbative series
in non-asymptotically free theories such as $\lambda\phi^4$ or quantum
electrodynamics \cite{Espriu:1996sk}. Second, it should be noted that the
stress-energy tensor defined as the variation of the effective action
is the consistent one. In fermionic theories it is also useful to
consider new stress-energy tensors obtained from the consistent one by
addition of local polynomial terms (in order not to modify its finite,
unambiguous, part) which do not necessarily come as the variation of
local polynomial terms in the effective action. This allows to obtain
the covariant stress-energy tensor, which is not consistent but it is
covariant under local Lorentz-frame transformations
\cite{Bardeen:1984pm}.

\section*{Acknowledgments}
This work is supported in part by funds provided by the Spanish DGICYT
grant no. PB95-1204 and Junta de Andaluc\'{\i}a grant no. FQM0225.


\begin{references}
\bibitem{note3}Note that the effective action itself needs not be a
polynomial or even analytical as a function of $m^2$, only the allowed
ambiguities should be polynomial. This allows for instance for terms
of the form $m^4\log(m^2/\mu^2)$ in the effective action, where $\mu$
is a mass scale introduced by the renormalization.
\bibitem{note4}For a coupling of the form
$L_\mu(x)\phi(x)\partial_\mu\phi(x)$, each new insertion of $L_\mu(x)$
decreases the degree of divergence and so the required counterterms
are still local polynomials in $L_\mu(x)$ of canonical dimension at
most four.
\bibitem{note5}Let us emphasize that this counting refers to the total number
of derivatives. Thus for instance, $R^n$ contains $2n$ derivatives.
\end{references}
\end{document}